# A 5.4 Gbps real time quantum random number generator with compact implementation

**JIE YANG,[1] JINLU LIU,[1] QI SU,[2] ZHENGYU LI,[3] FAN FAN,[1] BINGJIE XU,[1,*] AND HONG GUO[3]**

[1]*Science and Technology on Security Communication Laboratory, Institute of Southwestern Communication, Chengdu 610041, China*
[2]*State Key Laboratory of Cryptology, Beijing 100878, China*
[3]*State Key Laboratory of Advance Optical Communication Systems and Networks, Center for Computational Science & Engineering (CCSE) and Center for Quantum Information Technology, School of Electronics Engineering and Computer Science, Peking University, Beijing 100871, China*
*\*xbjpku@pku.edu.cn*

**Abstract:** We present a random number generation scheme based on measuring the phase fluctuations of a laser with a simple and compact experimental setup. A simple model is established to analyze the randomness and the simulation result based on this model fits well with the experiment data. After the analog to digital sampling and suitable randomness extraction integrated in the field programmable gate array, the final random bits are delivered to a PC, realizing a 5.4 Gbps real time quantum random number generation. The final random bit sequences have passed all the NIST and DIEHARD tests.

## 1. Introduction

Random numbers are of extreme importance for a wide range of applications in both commercial and scientific fields, such as numerical simulations [1], lottery games and cryptography [2]. A significant example is the quantum key distribution (QKD), in which the true random numbers are essential for both quantum state preparation and detection to guarantee unconditional security [3-5].

Classical pseudo random number generators (PRNGs) are based on the computational algorithms and have been widely used in modern digital electronic information systems. However, due to the deterministic and thus predictable features of the algorithms, PRNGs are not suitable for certain applications where true randomness is required [4,5]. Distinct from the PRNGs, true random number generators (TRNGs) rely on the randomness of physical processes. An important type of the TRNGs is the quantum random number generator (QRNG), which is based on the intrinsic randomness of fundamental quantum processes and can provide truly unpredictable and irreproducible random numbers [6].

Over the past two decades, various QRNG schemes have been proposed and demonstrated. For instance, one significant scheme based on the detection of single photon's path choice after a 50:50 beam splitter has been generally studied and implemented [7-9]. Commercial products employing this scheme have already appeared on the market such as the ID Quantis with the highest random bit generation rate of 16 Mbps [10]. Another approach based on single photon detection is to measure the arrival times of photons on a photodetector, which generates more than one bit per photon and hence achieves a higher generation rate [11]. However, limited by the performance of single photon detector (SPD), especially its dead time, the room of improvements for the ultimate speed of both schemes is not much.

To achieve higher generation rate, new schemes have been proposed and demonstrated, including measuring the vacuum field fluctuations [12,13] and quantum phase fluctuations [14-16]. The vacuum field fluctuation is a good quantum source to extract randomness while the signal is weak and a fast short-noise limited homodyne detector is required, which is a challenge. The quantum phase fluctuation of a laser results from the spontaneous emission and is thus inherently quantum mechanical. By operating near the threshold, the spontaneous emission is enhanced and the quantum phase fluctuations will be significant enough to be detected by conventional photodetectors, which is of higher speed and lower cost compared to the SPD. Based on this scheme, QRNGs with a generation rate of 500 Mbps, 6 Gbps and 68Gbps have been proposed and demonstrated successively and respectively [14-16].

In this paper, based on the scheme of quantum phase fluctuations, we propose a 5.4 Gbps real time QRNG with a simple and compact implementation, which employs only one BS and has the potential for practical applications. We derive the analytical expressions of the detected signals and perform the corresponding numerical simulation. The simulation result shows considerable agreement with the experimental data. After analog to digital converter (ADC) sampling and randomness extraction in real time, which is integrated in the field programmable gate array (FPGA), the final random bit sequences have passed all the NIST and DIEHARD tests.

## 2. Experimental setup and system model

Experimental setup of the proposed QRNG is shown in Fig.1. A DFB laser diode with a center wavelength of 1550.12nm is driven by a butterfly packaged laser diode driver to emit continuous-wave (CW) beams. The laser diode is operated around the threshold to maximize the quantum noise and thus to guarantee the randomness. The CW beams are split into two paths by a 2×2 50/50 polarization-maintaining beam splitter (BS). One of the output ports is directly coupled into a 1.8 GHz photo-detector (PD, MenloSystems, FPD310). And the other is coupled into the delay loop, which consists of the BS and a 4m delay line (DL). The beams coupled into the delay loop will be delayed for 20ns by the 4m DL for every circulation of the loop and hence 20N ns after N circulations.

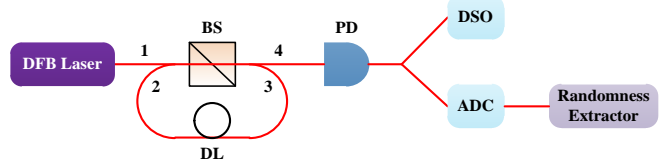

Fig. 1. The Experimental setup of the proposed QRNG. The CW beams emitted by the laser diode enter one input port (port 1) of the BS. Then half of the beams are directly coupled into the PD and another half of the beams are coupled into the delay loop, which consists of the BS and a 4m DL. Every circulation of the delay loop introduces a time delay of 20 ns. The beams from port2 will interfere with the beams from the laser diode at the BS. The output from the PD can be either acquired by a DSO to analyze the distribution of the raw data or be processed by an ADC and a randomness-extractor to distill the final random bits. CW: continuous wave, BS: 2×2 50/50 polarization-maintaining beam splitter, DL: delay line, PD: photo-detector with high bandwidth, DSO: digital storage oscilloscope, ADC: 12-bit analog-to-digital converter.

Theoretical model of the setup is established as follows. Suppose the field at port 1 of the BS is

$$E_{p1}(t) = A\exp\left[i\omega t + i\varphi(t)\right] \quad (1)$$

where $A$ is the amplitude of electric field, $\omega$ is the optical center angular frequency, and $\varphi(t)$ is the phase of the laser, respectively. The field at port 2 of the BS is the sum of beams that have transmitted in the delay loop for different times of circulations ranging from 1 to N (N → ∞), which includes all the components with significant values and can be calculated as

$$E_{p2}(t) = A\sum_{k=1}^{N}\left(\frac{1}{\sqrt{2}}\right)^k \exp\left[i\omega(t-k\Delta t) + i\varphi(t-k\Delta t)\right] \quad (2)$$

where $k \in [1, N]$ is an integer, and $\Delta t$ is the time delay induced by DL. $E_{p1}$ and $E_{p2}$ will interfere at the BS and the optical intensity, detected by the PD, can be given by

$$I = 1 + \sum_{k=1}^{n}\left(\frac{1}{2}\right)^k$$
$$+ \sum_{k=1}^{N}(\sqrt{2})^{2-k}\left(-\cos(k\omega\Delta t + \Delta\varphi_N^k)\right) \quad (3)$$
$$+ \sum_{k=1}^{N}(\sqrt{2})^{2-k}\sum_{j=1}^{k-1}\left(\frac{1}{\sqrt{2}}\right)^j \cos\left[(k-j)\omega\Delta t + \Delta\varphi_{N-j}^{k-j}\right]$$

where $j$ and $k$ are integers, and $k \in [1,N]$, $j \in [1,k)$, $\Delta\varphi_N^k = \varphi(N\Delta t) - \varphi((N-k)\Delta t)$ is the phase fluctuation between $N$-th order circulation and $N$-$k$-th circulation laser beam. Note that the DC component is $1 + \sum_{k=1}^{n}(1/2)^k$, which has no contribution to the generation of random bits. $\Delta\varphi_N^k$ is a Gaussian random variable due to spontaneous emission [17,18]. Therefore $I$, which is a superposition of $\Delta\varphi_N^k$, can be quantified to random bits.

To validate the proposed model, firstly a high performance DSO (Agilent, DSAV334A, 33 GHz bandwidth) is employed to acquire the experimental measured data, which maximally recovers the distribution of the raw output. Then based on the above theoretical model, the numerical simulation of the probability distribution of the output intensity is performed. In the experimental setup, the attenuation of the optical loop is 0.3dB, and the time delay $\Delta t = 20ns$, the line width of the laser working around the threshold level is about $5.5 MHz$ and hence $\Delta\varphi$ is modeled as $\Delta\varphi \sim N(0, 2\pi\Delta f \Delta t)$. The simulation result (red line) is shown in Fig.2, which fits well with the experimental measured data.

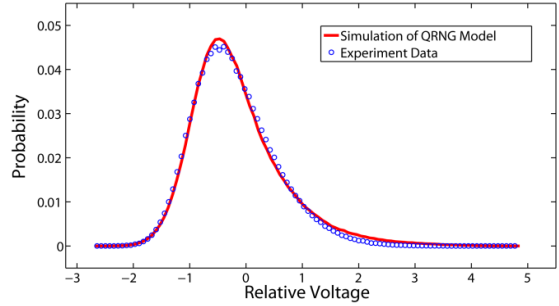

Fig. 2. The normalized distribution of the simulation result of the theoretical model for the proposed QRNG and the experimental measured interference intensity.

## 3. Randomness extraction and performance

To acquire the random bit sequence in real time and verify the randomness of the proposed QRNG, we employ a 12bit-ADC (TI, ADC12D1800, 1.8G SPS) to sample the raw output. And for hardware efficiency, the bitwise exclusive OR (XOR) operation and m-least-significant-bit (m-LSB) procedure are employed for randomness extraction, which is integrated in a field programmable gate array (FPGA) for real time processing.

The XOR operation is first performed on the raw data from the ADC for every 2 samples. Then we employ the min entropy, $H_{min} = -\log 2\{\max(p_i)\}$, as the indicator to decide how many bits can be reserved from each sample of the data after the XOR operation, which uses the maximal probability and represents the worst-case scenario. In our experiments, the typical value of the min entropy of the data after XOR operation is 9.592885, which means we can reserve at most 9 LSBs from each sample of the data after the XOR operation. To further illustrate the randomness extraction, we calculate the normalized min entropy $H_{nor-min} = H_{min}(l)/l$ of the random bit sequences after XOR operation and m-LSB procedure, where $H_{min}(l)$ is the min entropy calculated with the sample length of l, shown in Fig.3.

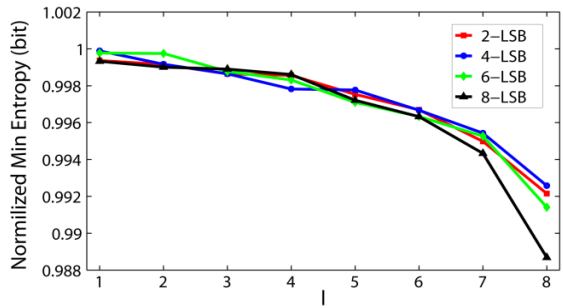

Fig. 3. Normalized min entropy $H_{min}(l)/l$ of $10^7$ extracted random bits after XOR operation and m-LSB procedure, $m = 2, 4, 6, 8$.

Since min entropy holds $H_{min}(l) = l$ for true random bit sequences, the normalized min entropy should be unity for any l. It can be seen from Fig. 3 that, after XOR operation and m-LSB procedure, the normalized min entropy is very close to unity.

Besides, Fig. 3 also shows that, when l is less than 7, there is no obvious difference between the normalized min entropies of the data for $m = 2, 4, 6, 8$. While when $l$ is 7 and 8, the normalized min entropies of the data for $m = 8$ falls lower than those of the data for $m = 2, 4, 6$. Therefore, comprehensively speaking, when $m$ is 2, 4 and 6, the final random bit sequences are of better randomness. So for the balance of randomness and random bit generation rate, and also for the consideration of the throughput capacity of the hardware interface, we reserve 6 LSBs from each sample after XOR operation to generate the final random bit sequences, yielding a 5.4 Gbps quantum random bit generation rate in real time.

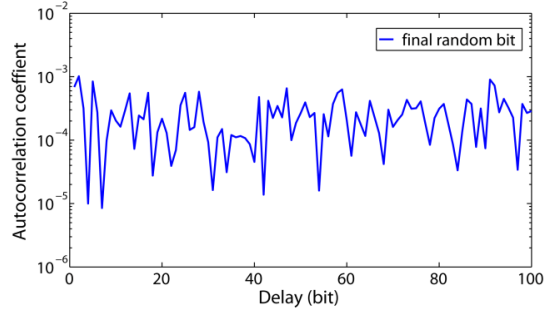

Fig.4. Autocorrelation analysis of $10^7$ extracted random bits after XOR operation and 6-LSB procedure within 100 bit-delay. The average value is $1.68 \times 10^{-5}$.

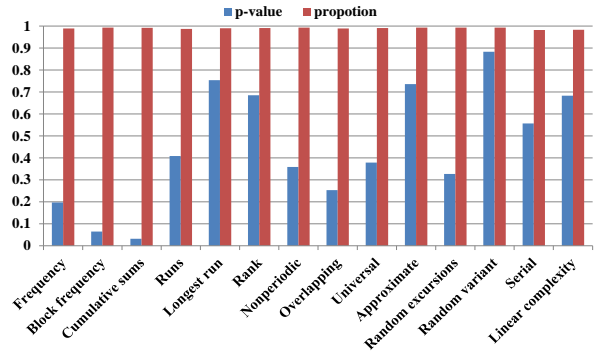

Fig.5. Results of the NIST-STS test suite for a 1Gbit sequence. The significance was set by 0.01. To past the test, p-value needs to satisfy $0.01 \leq p - value \leq 0.99$.

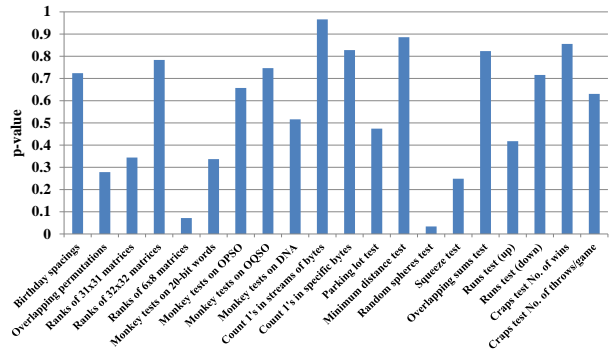

Fig.6. Results of the Diehard statistical suite for a 1Gbit sequence. For the case of multiple p-values in Diehard test suite, a Kolmogorov- Smirnov (KS) test is used to obtain a final p-value, which measures the uniformity of the multiple p-value.

In order to verify the randomness, firstly the autocorrelations within 100 bit-delay are calculated, shown in Fig.4, indicating good randomness. Then we applied two test batteries, the NIST-STS and Diehard, and the corresponding test results are shown in Fig.5 and Fig. 6. The random bits generated by the proposed QRNG can pass all the tests.

## 4. Conclusion

In this paper, a 5.4 Gbps real time QRNG is proposed based on the scheme of quantum phase fluctuations, which only employs one BS and therefore has the potential for practical applications. The analytical expressions of the detected signals are derived and the corresponding numerical simulation result shows considerable agreement with the experimental data. After ADC sampling and randomness extraction integrated in an FPGA, the final random bit sequences have passed all the NIST and DIEHARD tests.

**Acknowledgments**

This work is supported by the National Natural Science Foundation of China (Grant No. 61501414 and 61602045), the National Science Fund for Distinguished Young Scholars of China (Grant No. 61225003), the State Key Project of National Natural Science Foundation of China (Grant No. 61531003).